\documentstyle[aps,prb,psfig,floats]{revtex}
\begin{document}
\draft
\twocolumn[\hsize\textwidth\columnwidth\hsize\csname @twocolumnfalse\endcsname

\title{Epitaxial Growth Kinetics with Interacting Coherent Islands}

\author{H. M. Koduvely and A. Zangwill}

\address{School of Physics \\
         Georgia Institute of Technology \\
         Atlanta GA 30332-0430}
\date{\today}

\maketitle

\begin{abstract}
The Stranski-Krastanov growth kinetics of undislocated (coherent)
$3$-dimensional islands is studied with a self-consistent mean
field rate theory that takes account of elastic interactions
between the islands. The latter are presumed to facilitate the
detachment of atoms from the islands with a consequent decrease in
their average size. Semi-quantitative agreement with experiment is
found for the time evolution of the total island density and the
mean island size. When combined with scaling ideas, these results
provide a natural way to understand the often-observed initial
increase and subsequent decrease in the width of the coherent
island size distribution.
\end{abstract}
\vspace{2mm}
]
Heteroepitaxy begins with the formation of a thin, lattice-matched
wetting layer if the energy gain from substrate-adlayer adhesion
exceeds the elastic energy cost from lattice constant misfit
$\delta a/a$. As deposition continues, $2$-dimensional ($2D$)
islands nucleate on top of the wetting layer. These islands
contribute to the build-up of elastic strain and, for this reason,
the system does not tolerate their growth, coalescence, and
re-nucleation indefinitely. Instead, at large misfit, coherent
(undislocated) $3$-dimensional ($3D$) islands form that are lattice
matched near their base but are largely strain-relieved near their
top and sidewalls. Further deposition leads to their growth and
eventual coalescence. This is the so-called {\it
Stranski-Krastanov} growth mode \cite{review}.

A coherent island is the source of strain fields because it
elastically distorts the wetting layer and substrate in its
immediate vicinity. Early on, several experimental groups observed
a significant decrease in the mean size of the coherent islands at
relatively early stages of growth and suggested the possible role
of long range strain fields. For example, Ponchet and co-workers
\cite{ponchet1} presented data for the InAs/InP(001) system and
pointed out that elastic interactions should cause islands to {\it
destabilize} one another because their interactions are mutually
repulsive. Kobayashi and co-workers \cite{madhukar1} identified
several other features of the island-island interaction as a basis
for understanding their experiments on the InAs/GaAs(001) system.
Theoretical work on island interactions has been limited to {\it
equilibrium} considerations \cite{theory} up to the present time.

In this paper, we present a theoretical analysis of
Stranski-Krastanov growth kinetics that generalizes previous work
by Dobbs and co-workers \cite{dobbs} to take acccount of island
interactions and atom detachment from $3D$ islands. Dobbs {\it et
al.} employed a mean field theory for the density of adatoms $n_a$,
the density of $2D$ islands $n_2$, their average size $s_2$, the
density of $3D$ coherent islands $n_3$, and their average size
$s_3$. A rate equation was derived for each based on the physical
processes of adatom deposition, surface diffusion, attachment and
detachment of adatoms from the islands, etc. In brief, an incident
flux $F$ contributes directly to the increase of adatom popuation.
The adatoms diffuse on the surface with a diffusion constant $D =
\omega a^2
\exp(-E_s/k_B T)$ where $\omega$ is an attempt
frequency, $a$ is the lattice constant, $E_s$ is the energy barrier
for diffusion, $k_B$ is the Boltzmann constant and $T$ is the
temperature. Diffusing adatoms that meet bond together to form
small $2D$ islands but thermal fluctuations can cause them to break
apart if the island size is too small. There is a critical island
size $i$ such that islands of size $i$ and less are unstable.

An island grows by capturing adatoms from both the vapor and the
substrate. The rates for these processes are $F\kappa$ and $D
\sigma n_a$, where $\kappa$ is the direct capture number and
$\sigma$ is the diffusion capture number. To relieve strain, $2D$
islands convert into $3D$ islands at a rate $\gamma_2$. We assume
that atoms that detach from the edges of a $2D$ island do not leave
the island but instead migrate to the top of the island. On the
other hand, we suppose that atoms {\it do} detach from $3D$ islands
(at a rate $1/\tau_3$) when interactions become significant. A
fraction $m_2$ of these attach to $2D$ islands. The remaining
fraction $m_a$ contributes to the adatom population. In this work,
we approximate $m_2$ by the areal coverage of $2D$ islands.

Rate equations that incorporate all of these elementary processes
are
\begin{eqnarray}
\label{eq:twospe2}
\dot{n}_a~~ & = & F~[1 - (i+1)\kappa_i n_i - \kappa_2 n_2 -
\kappa_3 n_3] \nonumber  \\
&&  - D~[(i+1) \sigma_i n_i + \sigma_2 n_2 + \sigma_3
n_3]n_a \nonumber \\
&& + m_a n_3/\tau_3 \nonumber \\
& & \nonumber \\
\dot{n}_2~~ & = &
F~\kappa_i n_i + D~\sigma_i n_i n_a - \gamma_2 n_2  \nonumber \\
& &\nonumber \\
\dot{n}_3~~ & = & \gamma_2 n_2 \\
& & \nonumber\\
\dot{(s_2 n_2)}
&=& F [(i+1) \kappa_i n_i + \kappa_2 n_2] + D [(i + 1) \sigma_i n_i +
\nonumber \\
 && \sigma_2 n_2 ] n_a + m_2 n_3/\tau_3 - \gamma_2 s_2 n_2 \nonumber \\
& & \nonumber \\
\dot{(s_3 n_3)} &=& F \kappa_3 n_3 + D \sigma_3 n_3 n_a +
\gamma_2 s_2 n_2 - n_3/ \tau_3 \nonumber
\end{eqnarray}
The suffixes $i,2$ and $3$ for $\kappa$ and $\sigma$ denotes
critical nuclei (of size $i$), $2D$ and $3D$ islands respectively.

We assume that $2D$ islands are circular with radius $r$ and $3D$
islands are truncated pyramids with base length $l$, height $h$ and
base angle $\phi$. The radius of a $2D$ island is
$r=\sqrt{s_2/\pi}$. We assume that $3D$ islands very quickly
achieve their equilibruim shape and that the angle $\phi$ does not
change significantly during growth. For a given island size $s_3$,
$h$ and $l$ are found by minimizing the energy expression derived
for a $3D$ coherent island by Tersoff and Tromp \cite{tersoff1}.

The direct capture number $\kappa$ is given by the surface area of
the island normal to the incident flux, {\it i.e.}, $\kappa_2=s_2$
and $\kappa_3=l^2$. The diffusion capture number $\sigma$ measures
the efficiency with which an island captures adatoms from the
surface. We compute $\sigma$ using the prescription of Bales and
co-workers \cite{bales} which relates it to the size of the
diffusional depletion zone $\xi$ that surrounds each island. For a
circular island of radius $r$ with no barrier to adatom attachment
we solve self-consistently
\begin{eqnarray}
  \label{eq:sigma1}
  \sigma &=& 2 \pi \frac{ r K_1(r/\xi)}{\xi K_0(r/\xi)} \\
  \xi^{-1}    &=& \sqrt{F\kappa_i/D + (i+1) \sigma_i n_i +
  \sigma_2 n_2  + \sigma_3 n_3}
\end{eqnarray}
where $K_n(x)$ is the modified bessel function of order $n$. We use
equation (\ref{eq:sigma1}) for circular $2D$ islands and for
$\sigma_3$ as well (with $r$ replaced by $l/2$) because the details
of the island shape should not affect the results significantly.

Conversion of a $2D$ island to a $3D$ island occurs when a
sufficient density of atoms is present on its top (due to
strain-driven detachment from its perimeter and upward migration)
to nucleate a new island at its center.  The requisite nucleation
rate is \cite{dobbs}
\begin{equation}
  \label{eq:gamma2}
  \gamma_2 = \pi r^2 D \exp[(E_i - (i+1)E_d(r))/(k_B T)]
\end{equation}
where $E_i$ is the binding energy of critical nuclei and $E_d(r) =
E_0 ~ \ln(r/a)/(r/a)$ is a size dependent energy barrier for the
detachment of atoms from the $2D$ island. The form of $\gamma_2$ as
a function of $r$ is such that $2D$ islands barely convert at all
until they reach a size $s^*$ after which most of them convert very
rapidly.

The escape rate of an atom from a $3D$ island is
\begin{equation}
  \label{eq:escrate1}
  \frac{1}{\tau} = \frac{D}{a^2} \exp(-\frac{E_b}{k_BT})
\end{equation}
where $E_b$ is the energy barrier for detachment. Elasticity theory
\cite{dobbs} predicts that the change in the barrier due to strain
is $\Delta E_b
\approx (\eta_u
-\eta_s) \epsilon $, where $\epsilon$ is the local strain and
$\eta_s (\eta_u)$ is the local surface stress at the binding site
(transition state) configuration. The predicted linearity with
strain has been confirmed by first principles calculations
\cite{first}. The strain field due to a misfitting island is
proportional to the size of the island and varies as $d^{-3}$ for
distances $d$ far from the island \cite{srolovitz}. We therefore
put
\begin{equation}
  \label{eq:ebarri2}
  E_b = E_b^0 - 2\pi \alpha s_3 ({a \over d_3})^3
\end{equation}
where $E_b^0$ is the strain independent part of the energy barrier
and $d_3=1/\sqrt{n_3}$ is the average $3D$ island separation. We
treat $\alpha$ as an adjustable parameter because the surface
stress difference discussed above is difficult to estimate. The
factor of $2 \pi$ is, in this model, the mean number of islands
that are nearest-neighbors to a given island.
\begin{figure}[h]
  \begin{center}

    \psfig{figure=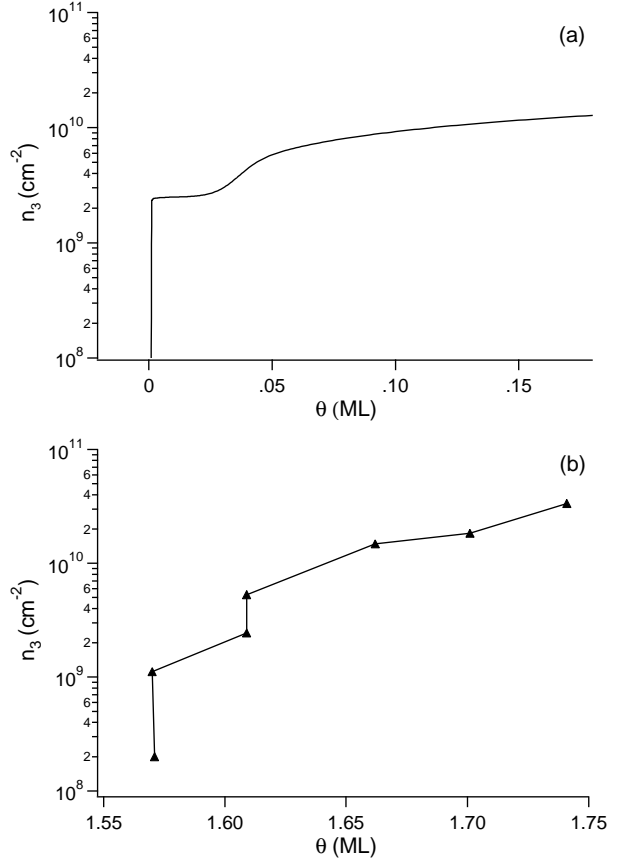,width=8.5cm}

    \caption{3D Island Densities:(a) from the present theory, (b) from
      the data of Kobayashi {\it et. al.} \cite{madhukar1}}
    \label{fig:3Dden}
  \end{center}
\end{figure}

\begin{figure}[h]
  \begin{center}

    \psfig{figure=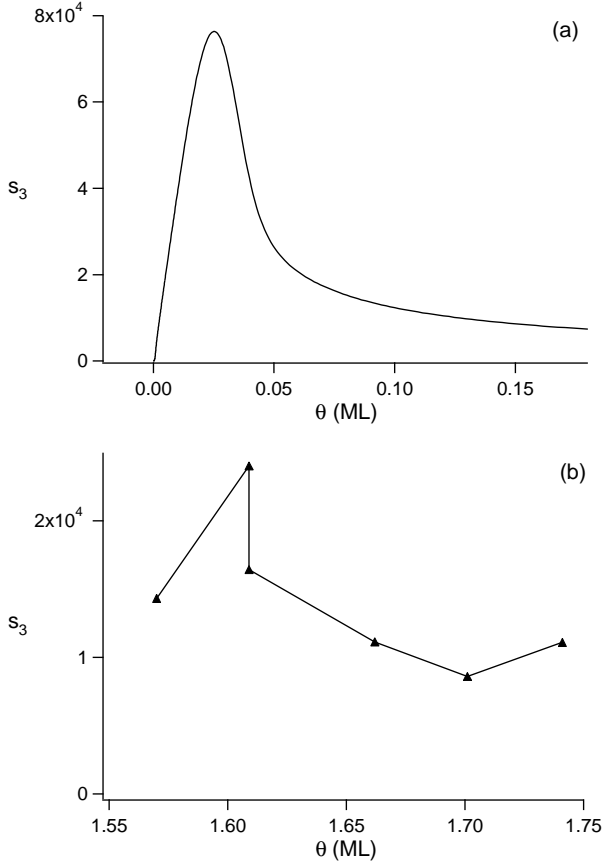,width=8.5cm}

    \caption{Average 3D Island Size: (a) from the present theory,
    (b) estimated from  the data of Kobayashi {\it et. al} \cite{madhukar1}}
    \label{fig:3Dsize}
  \end{center}
\end{figure}

The rate equations were integrated numerically using an algorithm
suited for systems of stiff differential equations \cite{bales}. We
used values of the parameters typical of those found in experiments,
$T= 900$ K, $F = 0.1$ ML/s, $a=3.0$ \AA, $i = 4$, $E_0=3.5$ ev, $E_i =
0.5$ ev, $E_s = 1.0$ ev, $E_b^0 = 0.7$ ev, $\phi = 25^0$ and $\delta a/a =
0.05$. Our results for the time (coverage) evolution of the $3D$
island density and mean size are shown in Figure \ref{fig:3Dden}(a)
and Figure \ref{fig:3Dsize}(a). For comparision we have plotted the
experimental results of Kobayashi {\it et al.} for InAs/GaAs(001)
\cite{madhukar1} in Figure \ref{fig:3Dden}(b) and
\ref{fig:3Dsize}(b). The sizes were estimated from the published
experimental distributions of island heights and island widths.  Note
also that we have shifted the theoretical curves to align the rapid
island density onsets because the precise onset position is related to
alloying \cite{joyce} that we do not attempt to model.

The $3D$ island density initially rises very rapidly due to the
fast conversion of $2D$ islands to $3D$ islands. It then tends to
saturate because, as a result of conversion, the average $2D$
island size decreases below $s^*$. During this time the average
$3D$ island size continue to grow. Soon the interactions become
important and significant detachment of atoms from the $3D$ islands
begins. This results in the very rapid decrease of $s_3$ seen in
Figure \ref{fig:3Dsize}(a). The detached adatoms that re-attach to
$2D$ islands increase the average size of the latter to $s^*$
which, in turn, leads to more $2D$ to $3D$ conversion. That is why
the $3D$ island density increases again. The same trend is seen in
the experimental data although we do not obtain quantitative
agreement between our model and the data.

The results shown correspond to $\alpha=120$ ev which is three
orders of magnitude greater than typical elastic energies. This
large number arises in our model because the rapid decrease in $3D$
island size seen in the data of Kobayashi {\it et al.}
\cite{madhukar1} occurs when the experimental mean island
separation is ten times larger than the mean island radius! Of
course, the real system has many islands at much closer distances
than our simple mean theory can describe, but it remains the case
that detachment effects seem to set in far earlier than simple
elasticity estimates would suggest. The detailed origin of this
behavior is an outstanding open question and our simple form
(\ref{eq:ebarri2}) must be regarded as a convenient
parameterization.

In principle, the entire island size distribution can be gotten
from a rate equation analysis. In practice however, it is
prohibitively difficult to solve the tens of thousands of equations
so generated. This theoretical problem is ameliorated for the case
of $2D$ {\it homoepitaxy} because the island size distribution
shows scaling behavior \cite{stroscio}. It is therefore highly
signficant that Ebiko {\it et. al} have shown that the $3D$
coherent island size distribution for the InAs/GaAs(001) system
also shows scaling \cite{ebiko}. Their data fits remarkably well to
an analytic scaling form suggested for $2D$ homoepitaxy
\cite{family1}. In detail, the number of islands of size $s$, $n_s$
takes the form
\begin{equation}
  \label{eq:scaling1}
  n_s = \frac{\theta_c}{\langle s \rangle^2} f(\frac{s}{\langle
  s \rangle}).
\end{equation}
where $\theta_c =\sum_s s n_s$ and
\begin{equation}
  \label{eq:scaling2}
  f(u) = 1.1 u \exp(-0.27u^{3.7}).
\end{equation}
It is surprising that an island distribution that works well for
$2D$ islands works equally well for $3D$ islands. Even more
puzzling is the fact that (\ref{eq:scaling2}) applies only to
situations where atom detachment from $2D$ islands is strictly
forbidden ($i=1$) whereas the coherent islands studied here shrink
precisely due to copious detachment.

\begin{figure}[h]
  \begin{center}

    \psfig{figure=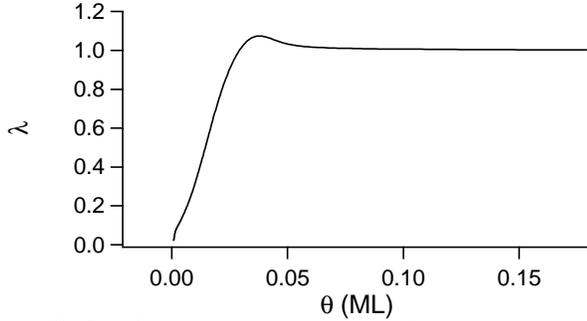,width=8.5cm}

    \caption{ Coverage dependence of $\lambda$,
     the detachment rate to attachment rate ratio for a 3D island.}
    \label{fig:lambda}

  \end{center}
\end{figure}

This can be understood if we parameterize the island size
distribution not by a fictitious "critical island size" but by the
ratio of the net detachment rate from an island to the net
attachment rate to an island \cite{evans}, namely
\begin{equation}
  \label{eq:lambda1}
  \lambda = \frac{1/\tau}{(F k + D \sigma n_a)}.
\end{equation}
Monte Carlo simulations of $2D$ homomepitaxy show that $\lambda$
parameterizes a continuous family of scaling functions
\cite{zang1}. When $\lambda \sim 1$ or less, the island size
distribution fits (\ref{eq:scaling1}) and (\ref{eq:scaling2}) very
well even when significant detachment is present. The computed time
evolution of $\lambda$ for our model is shown in Figure
\ref{fig:lambda}. Note that its value exceeds unity when the island
interactions are most important but only barely so. This is {\it
not} inconsistent with the rapid decrease in the average coherent
island size seen in Figure (\ref{fig:3Dsize}) because the rate
equation for this quantity in (1) involves the {\it difference}
(rather than the ratio) of the attachment and detachment rates
which can be large. These considerations provide a rationale for
the fitting procedure used by Ebiko {\it et al.} \cite{ebiko}.

\begin{figure}[h]
  \begin{center}

   \psfig{figure=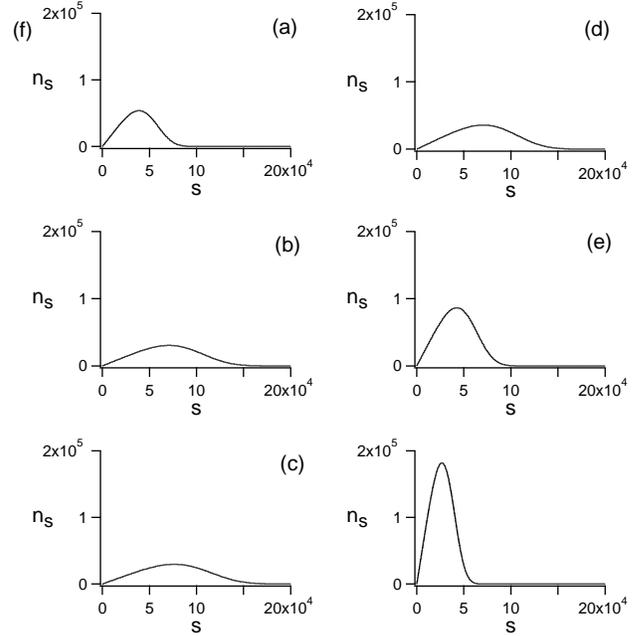,width=8.5cm}

    \caption{Evolution of the island size distribution: (1) $\theta =
    0.01$, (2) $\theta = 0.02$, (3) $\theta = 0.025$ (4) $\theta =
    0.03$, (5) $\theta = 0.04$, (6) $\theta = 0.05$ (ML)}
    \label{fig:distri1}

  \end{center}
\end{figure}
We conclude that island interactions strongly influence the average
island size but not the island size distribution scaling function.
This is important because it means that we can "synthesize" the
time dependence of the entire island size distribution merely from
knowledge of the time dependence of the average size. This is shown
in Figure (\ref{fig:distri1}). As expected, the island size
distribution broadens and its peak position moves to the right as
the coverage increases from zero. But as a consequence of equations
(\ref{eq:scaling1}) and (\ref{eq:scaling2}), the decrease of $s_3$
when interactions become imporant induces a narrowing of the
distribution and a shift back to the left. Precisely this behavior
is seen in the experimental island size distributions
\cite{ponchet1,madhukar1}.

In summary, we have generalized the theory of Dobbs {\it et. al.}
\cite{dobbs} to take account of island-island elastic interactions
that are presumed to induce atom detachment from $3D$ coherent
islands. Semi-quantitative agreement was found with experimental
results for InAs/GaAs(001) but the large value for the interaction
parameter needed to model the data suggests that we still lack a
good understanding of the energy barriers to detachment for this
problem. In conjunction with a scaling ansatz, the results could be
used nonetheless to rationalize the ubiquitous "narrowing" of the
full island size distribution seen in experiment. An interesting
and open question is to establish the veracity of this scaling
assumption in a theoretical framework.

The authors thank Steve Bales for the use of his program to solve
the rate equations. One of us (H.M.K.) gratefully acknowledges
financial support from NSF DMR 9705440.

\end{document}